\documentstyle[epsfig,amssymb,twocolumn]{mn2e}
\newif\ifAMStwofonts

\title[Gravitational drag]
{Gravitational drag on a point mass in hypersonic motion through a
gaseous medium}
\author[J. Cant\'o, A. C. Raga, A. Esquivel, F. J. S\'anchez-Salcedo]{
J. Cant\'o$^{1}$,
A. C. Raga$^{2}$\thanks{E-mail: raga@nucleares.unam.mx},
A. Esquivel$^{2}$,
F. J. S\'anchez-Salcedo$^{1}$\\
$^{1}$Instituto de Astronom\'\i a,
Universidad Nacional Aut\'onoma de M\'exico,
Ap. 70-468, 04510 D. F., M\'exico\\
$^{2}$Instituto de Ciencias Nucleares,
Universidad Nacional Aut\'onoma de M\'exico,
Ap. 70-543, 04510 D. F., M\'exico\\
}
\begin{document}

\date{}

\pagerange{\pageref{firstpage}--\pageref{lastpage}} \pubyear{2011}

\maketitle

\label{firstpage}

\begin{abstract}
We explore a ballistic orbit model to infer the gravitational drag force
on an accreting point mass $M$, such as a black hole, moving
at a hypersonic velocity $v_{0}$ through a gaseous environment of
density $\rho_{0}$. The streamlines blend in the flow past the body
and transfer momentum to it. The total drag force acting on the body,
including the nonlinear contribution of those streamlines with small
impact parameter that bend significantly and pass through a shock,
can be calculated by imposing conservation of momentum. In this
fully analytic approach, the ambiguity in the definition
of the lower cut-off distance $r_{\rm min}$  in calculations of the
effect of dynamical friction is removed. It turns out that
$r_{\rm min}=\sqrt{e}GM/2v_{0}^{2}$. Using spherical surfaces of control
of different sizes, we carry out a successful comparison between
the predicted drag force and the one obtained from a high resolution,
axisymmetric, isothermal flow simulation. We demonstrate that
ballistic models are reasonably successful in accounting for both
the accretion rate and the gravitational drag. 
\end{abstract}


\section{Introduction}

A body moving in a background medium loses momentum 
due to its gravitational interaction with its own gravitationally 
induced wake. This process is often referred to as dynamical friction.
Chandrasekhar (1943) estimated the dynamical friction on a 
massive particle passing through a homogeneous
and isotropic background of light stars.
In the case where the perturber moves in a gaseous medium,
the gravitational drag is traditionally inferred as the gravitational 
attraction between the perturber and its own wake. In this approach,
the density structure of the wake is derived in linear perturbation 
theory by assuming that the body produces a small perturbation
in the ambient medium (Dokuchaev 1964; Ruderman \& Spiegel 1971;
Just \& Kegel 1990; Ostriker 1999; Kim \& Kim 2007;
S\'anchez-Salcedo 2009; Namouni 2010).
For a perturber moving on a rectilinear orbit at constant velocity, 
the steady-state linear theory predicts that the drag force vanishes for 
subsonic perturbers, while it becomes similar to the collisionless drag 
force for supersonic bodies. Ostriker (1999) considered the linear-theory
drag as a time-dependent rather than steady state problem and arrived
at the following formula for the gravitational drag force,
\begin{equation}
 F_{g}= \frac{4\pi \rho_{0}G^{2}M^{2}}{v_{0}^{2}}\left\{ \begin{array}{ll}
\mbox{$\frac{1}{2}\ln \left(\frac{1+{\mathcal{M}}}{1-{\mathcal{M}}}\right)
-{\mathcal{M}}$} & \mbox{if ${\mathcal{M}}<1$};\\
 \mbox{$\frac{1}{2}\ln \left(1-{\mathcal{M}}^{-2}\right)+
\ln\left(\frac{v_{0}t}{r_{\rm min}}\right)$} & \mbox{if ${\mathcal{M}}>1$.} \end{array} \right.
\label{eq:ostriker99}
\end{equation}
The perturber of mass $M$, which moves at velocity $v_{0}$ 
and Mach number ${\mathcal{M}}$ in a rectilinear orbit through a 
homogeneous medium with density $\rho_{0}$ and sound speed 
$c_{0}$, is assumed to be formed at $t=0$. The minimum radius 
$r_{\rm min}$ is the typical size of the perturber.
This formula has enjoyed widespread theoretical application 
(Narayan 2000; Escala et al.~2004;
Kim 2007; Conroy \& Ostriker 2008; Villaver \& Livio 2009;
Tanaka \& Haiman 2009; Nejad-Asghar 2010; Chavarr\'{\i}a et al.~2010).
Because of the linear-theory assumption, the above equation is properly 
valid only at $r\gg R_{\rm BH}$ where $R_{\rm BH}$ is the Bondi-Hoyle 
radius ($R_{\rm BH}\equiv GM/[c_{0}^{2}(1+{\mathcal{M}}^{2})]$). Therefore, 
Equation (\ref{eq:ostriker99}) is strictly valid for extended perturbers with 
a softening radius much larger than the Bondi-Hoyle radius.
In fact, for extended perturbers, S\'anchez-Salcedo \& Brandenburg (1999) 
found good agreement between the gravitational drag in full hydrodynamical 
simulations and Ostriker's formula. In particular,
for Plummer perturbers with softening radius $r_{s}$ much larger 
than $R_{BH}$, they found that $r_{\rm min}=2.25r_{s}$. An extension of 
Ostriker's formula for extended bodies orbiting in a stratified gaseous sphere
was given in S\'anchez-Salcedo \& Brandenburg (2001).

 In the case of point-like perturbers, like massive black holes, it is 
 reasonable to assume that $r_{\rm min}$ should be of the order of a few 
 $R_{\rm BH}$, but a nonlinear analysis is required to fix the uncertainty in 
 the definition of $r_{\rm min}$. In adiabatic simulations of axisymmetric 
 accretion flows past a gravitating absorbing object, Shima et al.~(1985) 
 computed the drag by considering two contributions: the aerodynamic force, 
 which is due to the accretion of momentum over the body surface, and the 
 gravitational force on the perturber by its own wake. They found that
 the numerical results were consistent with the estimates in linear theory.

The problem of the gravitational drag on a point-mass particle
has revived new interest to estimate the timescale of the orbital
decay of massive black hole binary in the centre of galaxies. 
Escala et al. (2004) simulated the orbital decay  of a single black
hole moving initially on a circular orbit in an isothermal gaseous sphere. 
They found that the gravitational drag is less peaked at $1\leq {\mathcal{M}} <2$ 
than predicted by Ostriker's formula with $\ln v_{0}t/r_{\rm min}=3.1$.
Tanaka \& Haiman (2009) combined the prescriptions of Ostriker (1999)
and Escala et al. (2004) into a formula that is used as a prescription
of the gaseous drag on black holes in numerical simulations. 
In order to isolate
the physical reason of the failure of Ostriker's formula, Kim \& Kim (2009) 
and Kim (2010) carried out axisymmetrical simulations of a massive body 
in rectilinear orbit with different values of the strength of the gravitational
perturbation due to the body as measured by
\begin{equation}
{\mathcal{A}}=\frac{GM}{c_{0}^{2} r_{s}},
\end{equation}
where $r_{s}$ is the softening radius of the Plummer perturber.
They find that the functional form of the gravitational drag
is not so peaked as the linear theory predicts
and conclude that the discrepancy between the numerical and Ostriker 
results are most likely due to the nonlinear effect. It is important
to note that in the  simulations of Escala et al.~(2004), Kim \& Kim
(2009) and Kim (2010), the perturber  simply provides a smooth
gravitational potential and does not hold any absorbing 
surface. Without any absorbing inner boundary condition,
a hydrostatic envelope with front-back symmetry is formed near the perturber. 
Because of the front-back symmetry, this large envelope provides a negligible 
contribution to the gravitational drag force. 

However, it is well-known that
accretion is a crucial ingredient in point-like objects, such as black
holes or stars. Different boundary conditions in the high density region
of the wake are expected to change the gas dynamics near the perturber
and the strength of the gravitational drag (e.g., Fryxell et al. 1987;
Naiman et al.~2011). For example, Ruffert (1996) simulated a 3D
quasi-isothermal flow with an absorbing boundary surrounding the point-like
object. Moeckel \& Throop (2009) carried out similar simulations,
but then also included an accretion disk orbiting the point source.

The aim of this paper is to
describe the contribution of the nonlinear inner wake to the gravitational 
drag on hypersonic perturbers by using the ballistic orbit theory
(Bondi \& Hoyle 1944; Lyttleton 1972; Bisnovatyi-Kogan et al.~1979).
Whereas this theory (the so-called model of line-accretion) has been 
extensively used as a powerful framework
to describe the gravitational interaction between a moving massive body 
and the  surrounding gaseous medium in the context of supersonic 
Bondi-Hoyle-Lyttleton  accretion (e.g., Koide et al.~1991; Edgar 2004),
it has been traditionally ignored as a tool to quantify the gravitational drag.
In fact, all analytical studies about the gravitational drag in gaseous media
have been based on the linear perturbation theory, following on the analysis 
of Dokuchaev (1964), Ruderman \& Spiegel (1971) and Rephaeli \& 
Salpeter (1980). In this paper we develop the ballistic orbit theory to provide
analytical expressions of, not only the mass accretion rate, but also
the nonlinear drag force on a hypersonic compact body. These estimates will be
compared with numerical results of an axisymmetric isothermal hydrodynamical 
simulation.

\section{The free-streaming flow solution}

Let us consider the axisymmetric flow generated
by a point mass $M$ which moves hypersonically
at a constant velocity $v_0$ inside a homogeneous gaseous environment
(see also Bisnovatyi-Kogan et al.~1979).
Fig. 1 shows a schematic diagram illustrating the trajectory
of a fluid parcel in a frame of reference at rest with respect
to the point mass.

Because $v_0$ (the upstream environmental velocity, see Fig. 1)
is hypersonic, we neglect the pressure force and
consider the ballistic trajectory of the fluid parcels in the
gravitational potential of the point mass. As they
have a positive $E=v_0^2/2$ energy (per unit
mass), the trajectories of the fluid parcels are
hyperbolae of the form~:
\begin{equation}
r=\frac{\xi^2}{\xi_0\,(1+\cos\theta)+\xi \sin\theta}\,,
\label{r}
\end{equation}
where $\xi$ is the impact parameter of the fluid parcel (see
Fig. 1) and
\begin{equation}
\xi_0\equiv \frac{GM}{v_0^2}\,,
\label{xi0}
\end{equation}
with $G$ being the gravitational constant. For deriving
Eq. (\ref{r}) one has to consider a generic hyperbolic
trajectory of the form $r=p/[1+\epsilon\,\cos(\theta-\theta_0)]$,
and then impose the upstream boundary condition and the conserved
angular momentum $\xi v_0$ to determine the constants $p$, $\epsilon$
and $\theta_0$.

From Eq. (\ref{r}), we see that the streamline intercepts
the symmetry axis (i.e. $\theta=0$, see Fig. 1) at a position
\begin{equation}
x_0=\frac{\xi^2}{2\xi_0}\,,
\label{x0}
\end{equation}
downstream from the perturber.
The material will therefore pile up in
a narrow, downstream wake surrounding the symmetry axis, forming 
a dense column of gas.

From the equation for the streamlines (Eq. \ref{r}) one can 
calculate the velocity components of the free-streaming flow along the $x$ and $y$-axes~:
\begin{equation}
v_x=\frac{v_0}{\xi}\left(\xi+\xi_0\sin\theta\right)\,;\,\,\,
v_y=-\frac{v_0\xi_0}{\xi}\left(1+\cos\theta\right)\,.
\label{vxy}
\end{equation}
Assuming that the environment has a homogeneous density $\rho_0$ far
upstream from the source, it is possible to obtain the density
$\rho(x,y)$  of the free-streaming flow as a function of position:
\begin{equation}
\frac{\rho}{\rho_0}=\frac{\xi^3}{y\left[2\xi_0(r+x)+\xi y\right]}=
\frac{\xi^{2}}{y\left(2\xi-y\right)}\,,
\label{rho}
\end{equation}
with
\begin{equation}
r=\sqrt{x^2+y^2}\,;\,\,\,\,\,\xi=\frac{1}{2}\left[y+
\sqrt{y^2+4\xi_0(r+x)}\right]\,.
\label{rxi}
\end{equation}
Note that $x$ is the distance along the symmetry axis
and $y$ the cylindrical radius. It is simple to see that $\rho\geq \rho_{0}$.
Equation (\ref{rho}) is not valid in the shocked column of gas
near the positive $x$-axis.
The density enhancement in this approach is different from that derived
in linear theory which predicts zero-enhancement outside the Mach cone 
(e.g., Ostriker 1999). Koide et al.~(1991) found a good accordance between
the analytical solutions (\ref{vxy})-(\ref{rxi}) and the numerical solution even for
Mach numbers as low as $1.4$.

From Eq.~(\ref{vxy}) we see that at the point in which the streamlines intercept the symmetry
axis (i.e., for $\theta=0$), the flow velocity 
has an $x$-component $v_x (0)=v_0$ (identical to the far upstream
flow velocity and independent of the impact parameter $\xi$ of
the flow parcel) and a $y$-component $v_y (0)=-2v_0\xi_0/\xi$. This
latter component of the velocity will be thermalized in a shock
surrounding the downstream wake. We will assume that the
post-shock thermal energy is radiated away instantaneously.

\begin{figure}
\epsfig{file=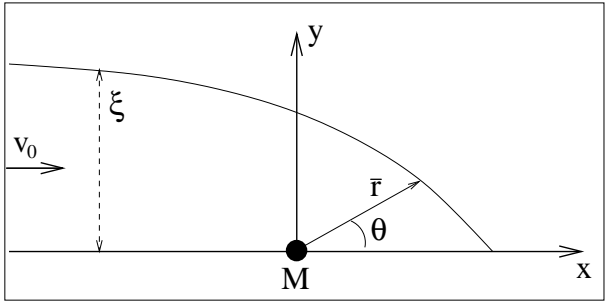,angle=0,width=8.0cm}
\caption{Schematic diagram showing the trajectory of an environmental
fluid parcel in hypersonic motion with respect to a point mass
$M$. The initial velocity $v_0$ of a parcel with impact parameter
$\xi$ is parallel to the $x$-axis, and its trajectory is the
$\overline{r}(\theta)$ curve. The problem has cylindrical symmetry,
with $x$ being the symmetry axis and $y$ the cylindrical radius.}
\label{f1}
\end{figure}

Now, the kinetic+potential energy per unit mass of the flow at $x\rightarrow
-\infty$ is
$E_0=v_0^2/2$. When the flow hits the symmetry axis (at $\theta=0$),
the energy associated with the $y$-velocity is thermalized, so that
the kinetic+potential energy is reduced to a value
\begin{equation}
E_t=E_0-\frac{v_y^2(0)}{2}=\frac{v_0^2}{2}\left[1-\left(\frac{2\xi_0}{\xi}
\right)^2\right]\,.
\label{et}
\end{equation}
From Eq. (\ref{et}) it is clear that $E_t\leq 0$ (i.e., the post-shock
material is gravitationally bound) if the condition
\begin{equation}
\xi\leq 2\xi_0
\label{xic}
\end{equation}
is met. Therefore all of the material arriving with impact
parameters $\leq 2\xi_0$ will eventually be accreted onto the body.
A streamline
with impact parameter $\xi_1=2\xi_0$ (with $\xi_0$ given by
Eq. \ref{xi0}) crosses the symmetry axis at a distance
\begin{equation}
x_1=2\xi_0\,,
\label{x1}
\end{equation}
downstream from the body (see Eq. \ref{x0} and Fig. 1).

The material within the downstream wake will have a complex
flow pattern. From Eq. (\ref{vxy}) it is clear that
the material enters the tail with a positive $x$-velocity.
The material with impact parameter $\xi\leq 2\xi_0$
(which is gravitationally bound, see above) will therefore
enter the wake flowing in the $+x$-direction, so that
it will first flow away from the body, and eventually
reverse and fall back onto the body. The distance $x_m$
from the body (along the $x$-axis) at which the flow
reverses can be obtained from the condition of zero
velocity for a radial motion in the gravitational
potential. This condition gives~:
\begin{equation}
x_m=\frac{2\xi_0}{(2\xi_0/\xi)^2-1}\,.
\label{xm}
\end{equation}
We see that when $\xi\rightarrow \xi_{1}$, $x_{m}\rightarrow \infty$.
Consequently, streamlines with impact parameter close to and
smaller than $\xi_{1}$, will take a long time to be accreted.

The material in the wake that remains gravitationally
unbound when entering the wake (i.e., the material
with impact parameters $\xi>2\xi_0$, see above), will
flow away from the body along the $x$-axis, reaching infinity
with a velocity
\begin{equation}
v_\infty=v_0\left[1-\left(\frac{2\xi_0}{\xi}\right)^2\right]^{1/2}\,.
\label{vinf}
\end{equation}

\section{The mass accretion rate and the drag force}

\subsection{The accretion rate}

From the solution of section 2, the accretion rate onto the
point mass can be obtained (Hoyle \& Lyttleton 1939;
Bondi \& Hoyle 1944). All of the
material with impact parameters $\xi\leq 2\xi_0$ (see eq. \ref{xic})
will fall onto the body. Therefore, the mass accretion rate is:
\begin{equation}
{\dot M}_{acc}=\pi (2\xi_0)^2 \rho_0 v_0=\frac{4\pi(GM)^2\rho_0}{v_0^3}\,,
\label{m}
\end{equation}
where we have used Eq. (\ref{xi0}) for the second equality.

\subsection{The gravitational drag}

We calculate the gravitational drag on the perturber by computing the
net $x$-momentum per unit of time, ${\dot \Pi}_x$, going through a spherical control
volume of radius $R>2\xi_0$ centred on the body. It is clear that the
contribution to the drag by the gas within the sphere is equal to ${\dot \Pi}_x$.

We consider the three streamlines shown in the schematic diagram
of Fig. 2~:
\begin{itemize}
\item a streamline with impact parameter $\xi_1=2\xi_0$, which
crosses the axis at $x_1=2\xi_0$ (see eq. \ref{x1}). All of
the material with $\xi\leq \xi_1$ is accreted onto the body,
\item a streamline with impact parameter $\xi_2=\sqrt{2\xi_0 R}$,
which crosses the axis at a distance $R$ downstream from the perturber
(where $R$ is the radius of the control sphere, see above),
\item a streamline with impact parameter $\xi_3$, which
tangentially touches the control sphere (at a point
with polar angle $\theta_3$, see Fig. 2).
\end{itemize}

In order to obtain $\xi_3$ and $\theta_3$ we first set $r=R$ (i.e., a
radius equal to the radius of the control sphere) in Eq. (\ref{r}),
and invert this equation to find~:
\begin{equation}
\sin\theta_\pm=\frac{\xi\left[\xi^2-\xi_0R\pm\xi_0
\sqrt{R(2\xi_0+R)-\xi^2}\right]}{R(\xi^2+\xi_0^2)}\,,
\label{th}
\end{equation}
which gives the two values of $\theta$ at which the streamline
with impact parameter $\xi$ cuts the control sphere. The
angle $\theta_+$ (obtained with the $+$ sign of the right
hand term of Eq. \ref{th}) corresponds to the point in which
the streamline enters the control sphere, and $\theta_-$ corresponds
to the exit point. For the tangential
streamline (with impact parameter $\xi_3$, see Fig. 2) the entry
and exit points coincide, so that the term within the square root
of Eq. (\ref{th}) has to be equal to zero. From this condition,
we obtain the value of $\xi_3$~:
\begin{equation}
\xi_3=\sqrt{R(2\xi_0+R)}\,,
\label{xi3}
\end{equation}
and using Eq. (\ref{th}) we then obtain
\begin{equation}
\sin\theta_3=\frac{\sqrt{R(2\xi_0+R)}}{R+\xi_0}\,.
\label{th3}
\end{equation}

Now, the $x$-momentum entering the control sphere from the
upstream region can be calculated as~:
\begin{equation}
{\dot \Pi}_{x,in}(R)=2\pi\rho_0 v_0\int_0^{\xi_3}v_x(\xi,\theta_+)\,\xi d\xi\,,
\label{pin}
\end{equation}
where $\xi_3$ is given by Eq. (\ref{xi3}), $v_x(\xi,\theta)$ by
Eq. (\ref{vxy}) and $\theta_+$ is obtained with the $+$ sign
of Eq. (\ref{th}). This integral can be solved analytically to
obtain~:
\begin{equation}
{\dot \Pi}_{x,in}(R)=4\pi \xi_0^2\rho_0 v_0^2\,f_{in}\,,
\label{pin1}
\end{equation}
where
$$f_{in}=\frac{1+3w_0}{4w_0^2}-$$
\begin{equation}
\frac{1}{2}\left[\sqrt{1+2w_0}-1+(1+w_0)
\ln\left(\frac{1+w_0}{1+w_0+\sqrt{1+2w_0}}\right)\right]\,,
\label{fin}
\end{equation}
with $\omega_{0}\equiv \xi_{0}/R<1/2$.

The $x$-momentum rate leaving the control domain has two terms:
\begin{itemize}
\item the rate ${\dot \Pi}_{x,b}$ of $x$-momentum leaving through the
boundary of the spherical domain,
\item the rate ${\dot \Pi}_{x,a}$ of $x$-momentum hitting the symmetry
axis and exiting the domain through a narrow wake along the $x$-axis.
\end{itemize}

The momentum rate leaving the sphere through the boundary of the control volume
is given by~:
\begin{equation}
{\dot \Pi}_{x,b}=2\pi\rho_0 v_0\int_{\xi_2}^{\xi_3}
v_x(\xi,\theta_-)\,\xi d\xi\,.
\label{pob}
\end{equation}
This integral can be performed analytically to obtain~:
\begin{equation}
{\dot \Pi}_{x,b}=4\pi \xi_0^2\rho_0 v_0^2\,f_b\,,
\label{pob1}
\end{equation}
where
\begin{equation}
f_b=\frac{1+w_0}{4w_0^2}+
\frac{1}{2}
\left[1+(1+w_0)
\ln\left(\frac{w_0}{1+w_0}\right)\right]\,.
\label{fb}
\end{equation}

The momentum rate exiting the control region through the wake is
given by~:
\begin{equation}
{\dot \Pi}_{x,a}=\int_{\xi_1}^{\xi_2} v_R\,d{\dot m}\,,
\label{poa}
\end{equation}
where
\begin{equation}
d{\dot m}=2\pi \xi \rho_0 v_0d\xi\,,
\label{dm}
\end{equation}
and the velocity $v_R$ along the axis with which the material
leaves the control domain is given by the kinetic+potential
energy conservation condition
\begin{equation}
\frac{v_0^2}{2}-\frac{GM}{x_0}=\frac{v_R^2}{2}-\frac{GM}{R}
\label{vr}
\end{equation}
where $x_0$ is the distance along the $x$-axis 
at which the streamline intercepts the axis as given by
Eq. (\ref{x0}). Here we have used that the $x$-component of
the velocity is $v_{0}$, as derived in Eq.~(\ref{vxy}).
Using Eqs. (\ref{dm}-\ref{vr}), the integral in Eq. (\ref{poa})
can be performed analytically to obtain~:
\begin{equation}
{\dot \Pi}_{x,a}=4\pi \xi_0^2\rho_0 v_0^2\,f_a\,,
\label{poa1}
\end{equation}
where
$$f_a=\frac{1-(2w_0)^{3/2}}{2w_0}+$$
\begin{equation}
\frac{1}{\sqrt{1+2w_0}}
\ln \left[\frac{2w_0+\sqrt{2w_0(1+2w_0)}}{1+\sqrt{1+2w_0}}\right]\,.
\label{fa}
\end{equation}

Finally, the net drag force of the gas on the body
is obtained as~:
\begin{equation}
F_d={\dot \Pi}_{x,in}-{\dot \Pi}_{x,b}-{\dot \Pi}_{x,a}=
\frac{4\pi (GM)^{2}\rho_{0}}{v_{0}^{2}}\,f(\xi_0,R)\,,
\label{fd}
\end{equation}
where $\xi_0$ is given by Eq. (\ref{xi0}) and
\begin{equation}
f(\xi_0,R)=f_{in}-f_b-f_a\,,
\label{f}
\end{equation}
with $f_{in}$, $f_b$ and $f_a$ given by Eqs. (\ref{fin}),
(\ref{fb}) and (\ref{fa}), respectively.
Note that Equation (\ref{fd}) includes the force on the body due to 
momentum accretion (sometimes called as aerodynamic force).

It is straightforward to show that in the $R\gg \xi_0$ limit
($w_0\ll 1$, see Eqs. \ref{fin}, \ref{fb} and \ref{fa}), this
function takes the form
\begin{equation}
f(\xi_0,R)\approx \ln\left(\frac{2R}{\xi_0}\right)-\frac{1}{2}
\left(1-\frac{\xi_0}{R}\right)\,.
\label{fp}
\end{equation}
A comparison between the full (Eq. \ref{f}) and approximate
(Eq. \ref{fp}) forms of $f$ is shown in Fig. 3. It is clear
that for radii larger than $\sim 5\xi_0$ the two forms of
$f$ agree to better than $\sim 10$\%\ .

From Eq. (\ref{fp}) we see that the drag force diverges
logarithmically for large values of $R$, as occurs in 
linear perturbation theory. Therefore, it is possible to
match the solution found in linear theory (Eq.~\ref{eq:ostriker99}), which is
valid at far enough distances from the body, with that found in the nonlinear
analysis. This can be accomplished by replacing
$R$ for $v_{0}t$, where $t=0$ is the time at which
the body is formed\footnote{In a realistic situation, $R$ will
increase with time until it reaches the boundary of the cloud.}.
Moreover, we see that the ambiguity in the definition of the
minimum radius $r_{\rm min}$ that appears in linear theory
is removed in our framework. In fact, we find that 
$r_{\rm min}=\sqrt{e}\xi_{0}/2$, where the factor $\sqrt{e}$ comes
from inserting the term $-1/2$ that appears in the right-hand-side 
of Equation (\ref{fp}) in the argument of the log.

\begin{figure}
\centering
\epsfig{file=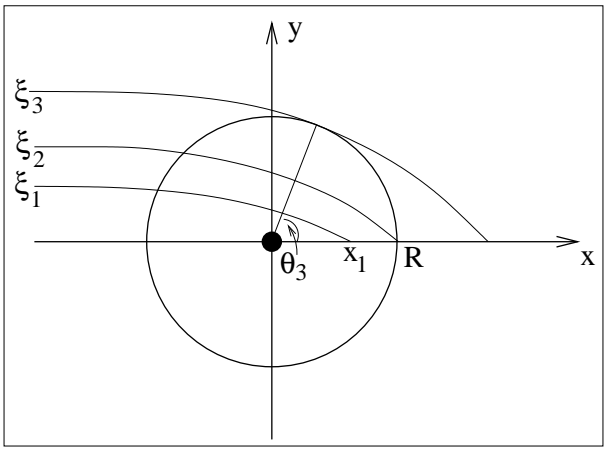,angle=0,width=8.0cm}
\caption{Schematic diagram showing the control sphere (of radius
$R>x_1=2\xi_0$) and the three streamlines used in the calculation of
the drag force (see section 3.2).}
\label{f2}
\end{figure}

\begin{figure}
\centering
\epsfig{file=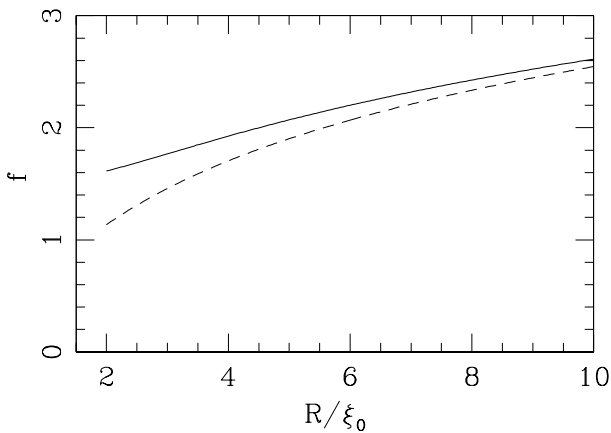,angle=0,width=8.0cm}
\caption{Exact (solid line, see Eq. \ref{f}) and approximate
(dashed line, see Eq. \ref{fp}) forms of the $f$ function,
which gives the dependence of the drag force as
a function of the radius $R$ of the control volume.}
\label{f3}
\end{figure}

\section{An axisymmetric numerical simulation}

We have computed an axisymmetric numerical simulation, solving
the Euler equations for an isothermal flow in a uniform, cylindrical
computational grid. We have used the ``flux vector splitting''
algorithm of van Leer (1982), with the second order (time and space)
implementation described by Raga et al. (2000).

The simulation that we are presented can be compared with the work
of Ruffert (1996) and Moeckel \& Throop (2009), who computed
3D simulations of basically the same physical situation. The main
difference with this previous work is that our simulation is 2D
(axisymmetric), and has $\sim 2$ orders of magnitude higher
resolution.

The computational domain has an axial extent of $15\xi_0$ (with
$\xi_0$ being the gravitational radius given by Eq. \ref{xi0})
and a radial extent of $7.5\xi_0$, resolved with $9000\times 4500$
(axial $\times$ radial) grid points. A point mass (influencing
the flow only through its gravitational attraction) is placed
in the middle of the axial extent of the domain. A spherical
volume of radius $0.05 \xi_0$ (30 pixels)
around the body is artificially
kept at a low density at all times, so that the material entering
this volume from the rest of the computational domain is effectively
removed. We simulate in this way the accretion of gas onto the
object.

In the left boundary of the domain we impose an inflow of
density $\rho_0$ and velocity $v_0$, parallel to the symmetry
axis. A reflection condition is applied on the symmetry axis,
and a zero gradient condition is applied in the remaining
two boundaries of the computational domain. In the initial
condition, the domain is filled with a uniform flow (of
velocity $v_0$ and density $\rho_0$) parallel to the symmetry
axis. The isothermal sound speed is chosen to be $c_0=v_0/5$
(i.e., the flow entering the domain has a Mach number of 5).

The results obtained after time-integrations of 10, 40 and $70\xi_0/v_0$
are shown in Fig. 4. This figure
is a zoom of an inner region of the computational domain, showing
the highly time-dependent wake formed downstream of the body.

\begin{figure*}
\centering
\epsfig{file=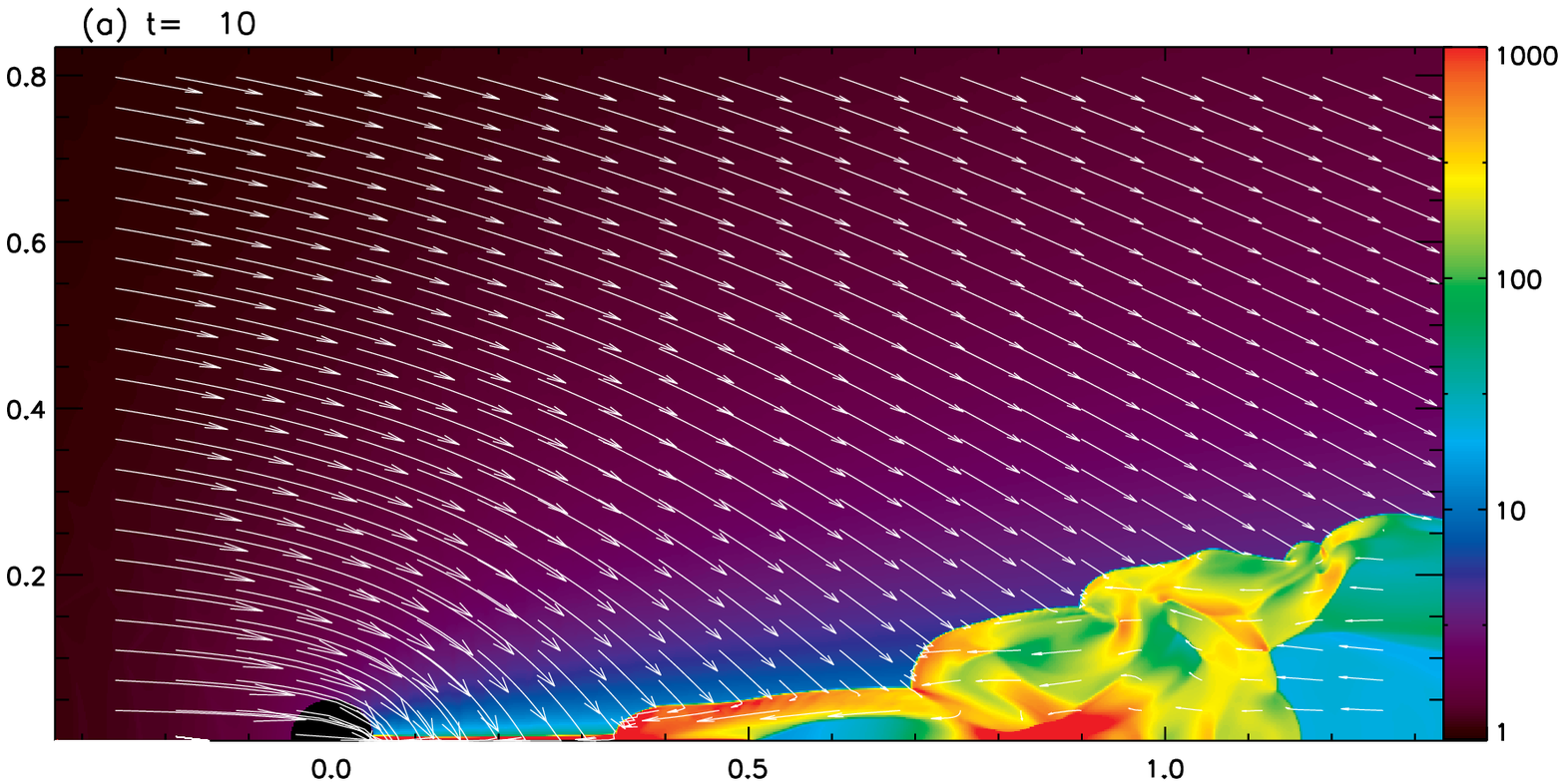,angle=0,width=15.0cm}
\epsfig{file=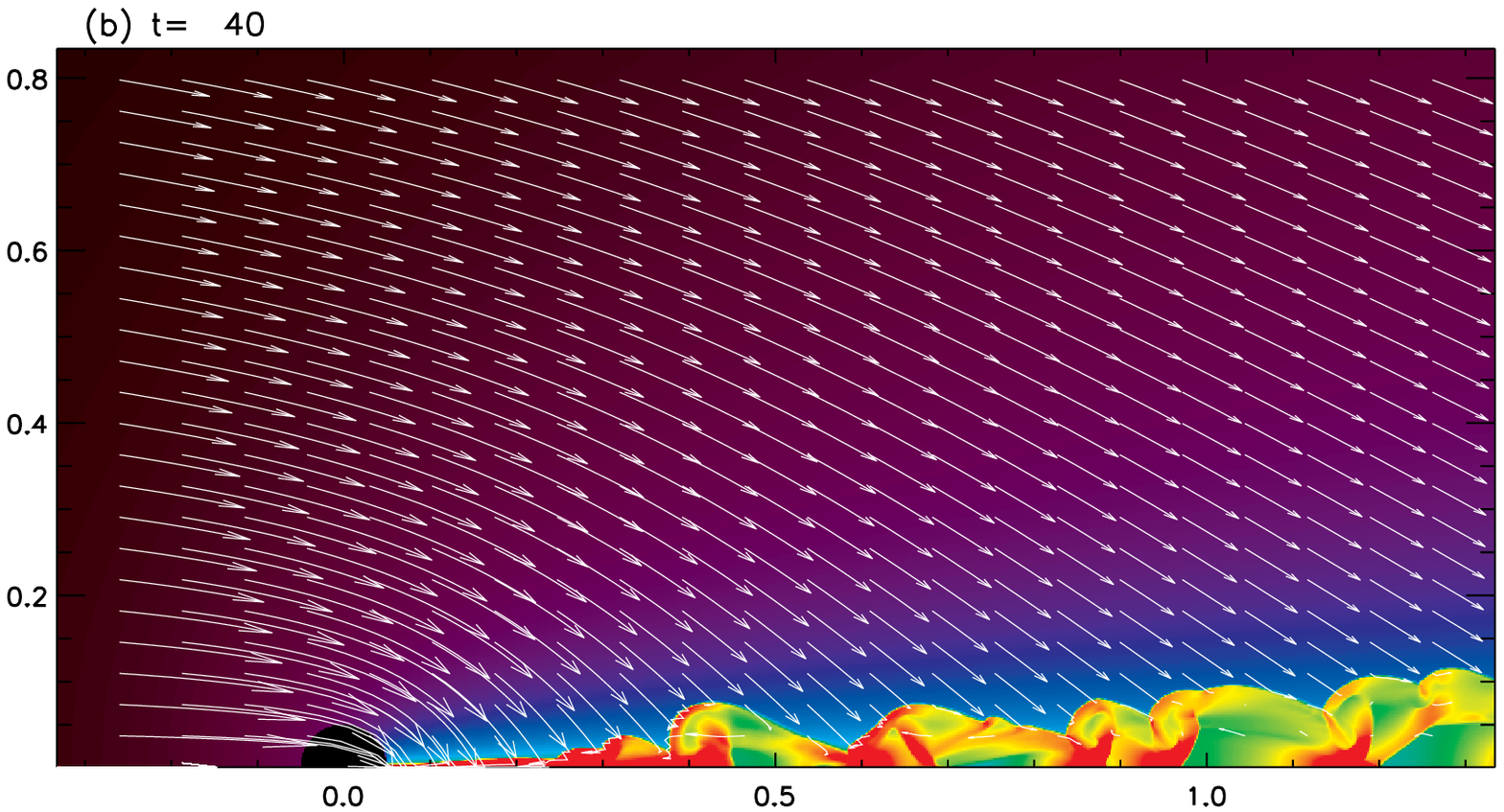,angle=0,width=15.0cm}
\epsfig{file=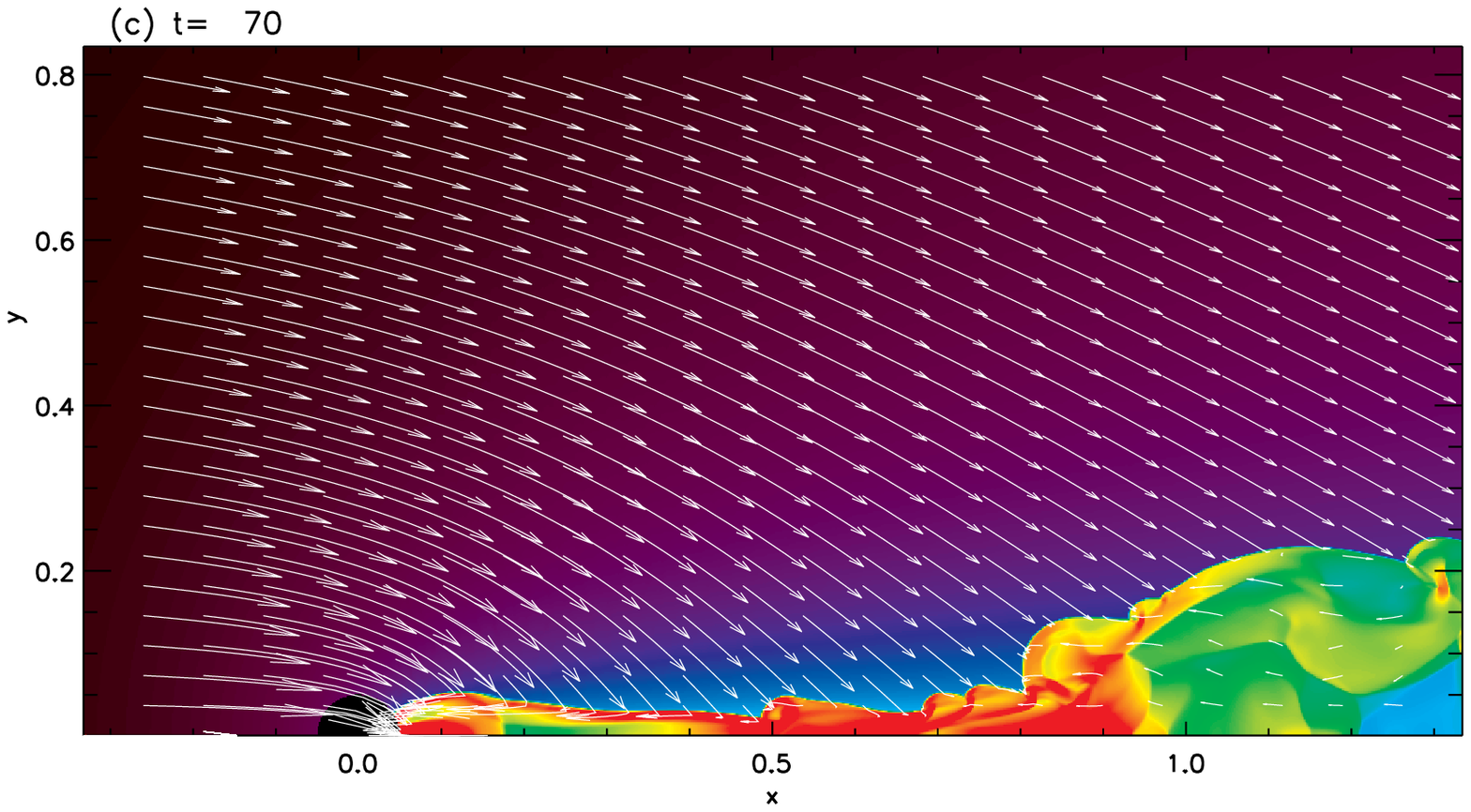,angle=0,width=15.0cm}
\caption{Density (in units of $\rho_0$, with the colour
scale given by the top right bar) and velocity field
(white arrows) from the axisymmetric simulation described in section
4, obtained for integration times $t=10\xi_0/v_0$ (panel $a$),
$40\xi_0/v_0$ (panel $b$) and $70\xi_0/v_0$ (panel $c$). The
axes are in units of $\xi_0$. The perturber is on the abscissa
at position $x=0$, and the flow enters the domain from the left.
Only a limited region of the computational domain is shown (see
section 4). The $x$ (symmetry axis) and $y$ (cylindrical radius)
axes are labeled in units of $\xi_0$.}
\label{f4}
\end{figure*}

We take the density and flow velocity time-frames obtained from
the simulation, and compute the net mass ${\dot M}_{acc}$, 
and momentum fluxes
through a control sphere of arbitrary radius $R$ centred on the
point mass.
Thereby, only ${\dot M}_{acc}$ computed with $R=2\xi_0$
corresponds exactly to the mass accretion rate onto the body.
 We also compute the gravitational force exerted on the body
by the material within the control volume. In Fig. 5, we show
the mass flux ${\dot M}_{acc}$, the drag force $F_d$ and
the gravitational force $F_g$ computed with a control volume
of radius $R=5\xi_0$. $F_g$ is inferred as the gravitational
attraction between the body and the perturbed medium.
${\dot M}_{acc}$ and $F_d$ show a peak
at $t\approx 5 \xi_0/v_0$, and have fluctuating values
for $t\geq 10 \xi_0/v_0$. The gravitational force $F_g$
initially grows as more material enters the wake behind
the object, and also shows fluctuating values as a function
of time. The fact that ${\dot M}_{acc}$, $F_d$ and $F_g$
have strong fluctuations is not surprising given the
strongly time-dependent structure of the flow (see Fig.
4).
 
\begin{figure}
\centering
\epsfig{file=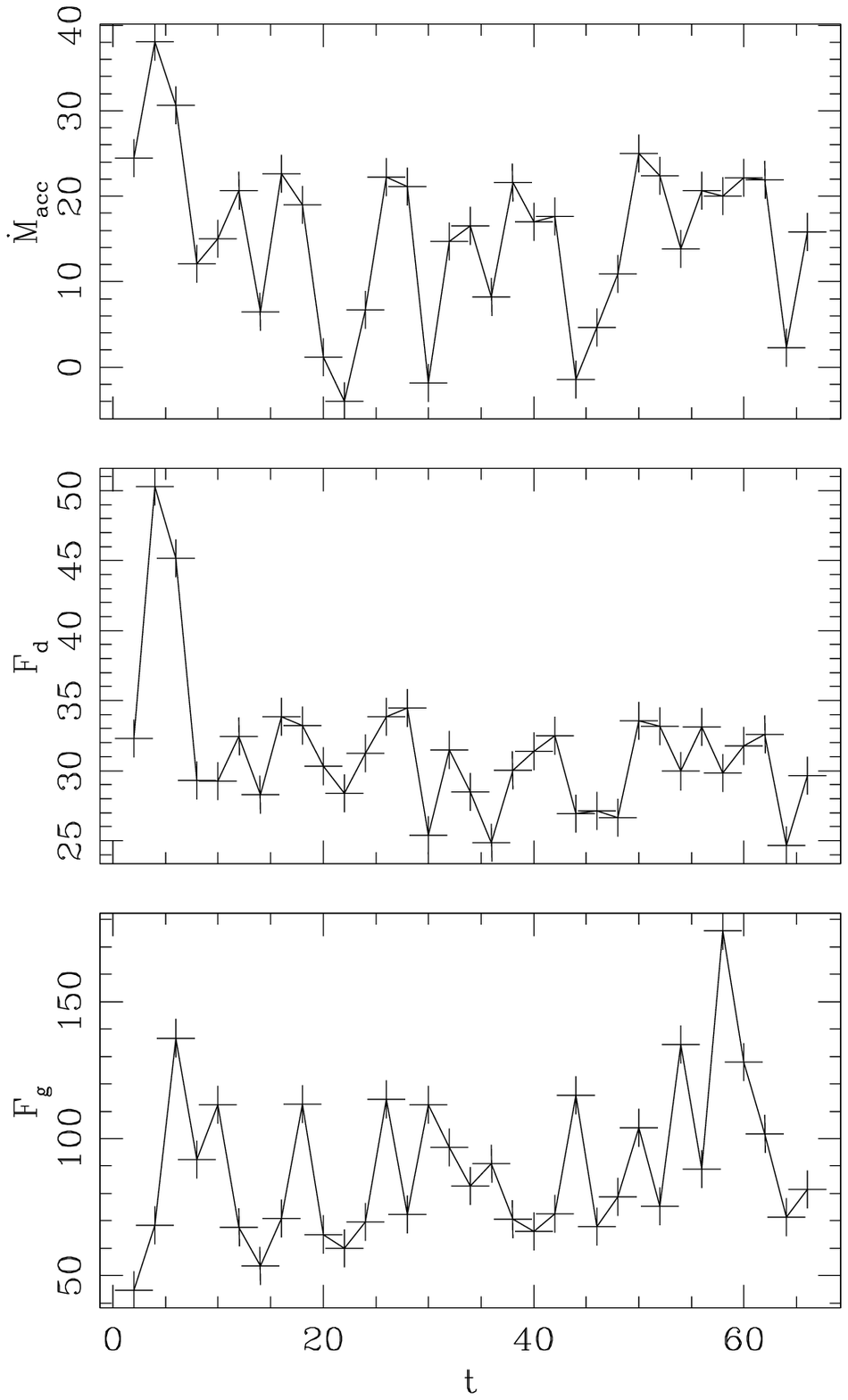,angle=0,width=8.0cm}
\caption{Mass flux (in units of $\xi_0^2 \rho_0 v_0$, top),
drag force (in units of $\xi_0^2 \rho_0 v_0^2$, centre)
and gravitational force on the body (in units of
$\xi_0^2 \rho_0 v_0^2$, bottom) as a function of time
(in units of $\xi_0/v_0$). These parameters were computed from the
results of the axisymmetric simulation (described in the
text) using a spherical control volume of radius $R=5\xi_0$.}
\label{f5}
\end{figure}

\begin{figure}
\centering
\epsfig{file=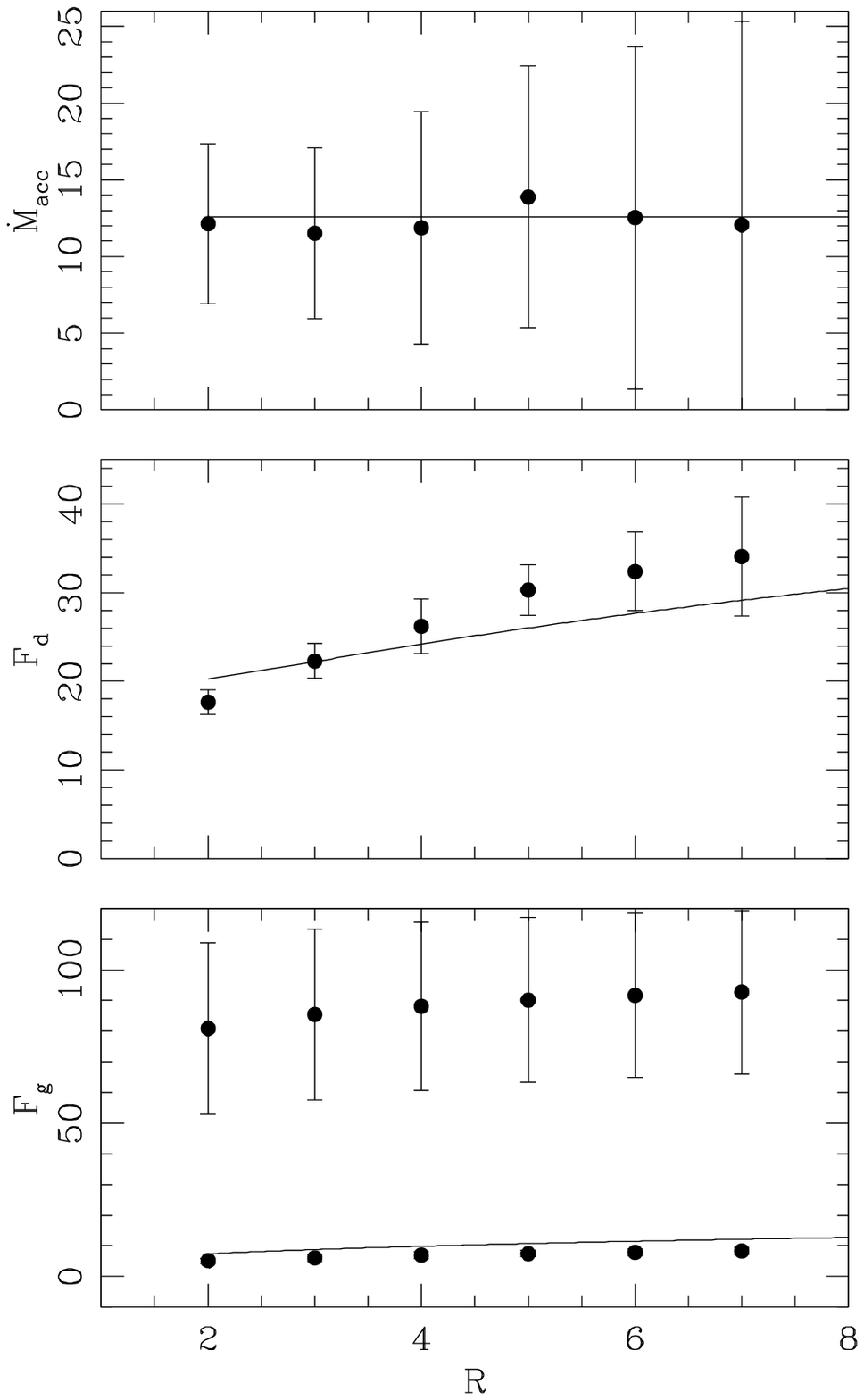,angle=0,width=8.0cm}
\caption{Time-averaged values of the net mass flux
(in units of $\xi_0^2 \rho_0 v_0$, top), drag force (in units
of $\xi_0^2 \rho_0 v_0^2=(GM)^{2}\rho_{0}/v_{0}^{2}$, centre)
and gravitational force on the body (in the same units, bottom)
as a function of the radius $R$
of the control volume ($R$ is given in units of $\xi_0$). The
dispersions are indicated by the error bars. The gravitational
force has been calculated in two ways: considering the contribution
of all of the gas within the control sphere (sequence of points
on the upper part of the bottom frame), and eliminating the
contribution of the material within the wake (lower sequence
of points, bottom frame). The solid lines
are the predictions from the analytic model described in section
3.}
\label{f6}
\end{figure}

In order to carry out a comparison with the analytic model
(see section 3.2), we have calculated the average
values and the dispersions of ${\dot M}_{acc}$, $F_d$ and
$F_g$ in the interval $10\xi_0/v_0\leq t \leq 66\xi_0/v_0$,
in which the fluctuations of these quantities appear
to be statistically stationary (see Fig. 5). We then
plot these time-averaged values as a function of the
radius $R$ of the control volume in Fig. 6.

We have considered control volumes with radii $2\xi_0\leq R
\leq 7\xi_0$, the lower boundary being fixed by the derivation
of the analytic model (in which it was assumed that $R\geq 2\xi_0$,
see section 3) and the upper boundary given by the approach to the
outer edge of the computational grid. It is clear from Fig. 6 that
the dispersions of the ${\dot M}_{acc}$ and $F_d$ values grow
as a function of $R$ (due to the fact that larger, more massive
eddies are seen at larger distances downstream from the body,
see Fig. 4), while the dispersion of $F_g$ (which is a quantity integrated
over the volume of the control sphere) remains approximately
constant.

The analytic model predicts that ${\dot M}_{acc}=4\pi\xi_0^2\rho_0v_0$
(see Eq. \ref{m}) for all control spheres with $R\geq 2\xi_0$.
The top panel of Fig. 6 shows that the time-averaged values obtained
from the numerical simulation closely reproduce this result.
The central panel of Fig. 6 shows the drag force $F_d$ calculated
using Eq. (\ref{fd}), which has values that differ from the
results from the numerical simulations by less than $\sim 15$~\%\
(though the slope of the $F_d$ vs. $R$ dependence appears to be
higher in the numerical results than in the analytic model).

In the bottom panel of Fig. 6 we show the gravitational force on the body
due to the density structure of the
numerical model. The gravitational force is larger than
the net drag because momentum accretion onto the body produces
an accelerating force. 
We also plot the gravitational force from the
numerical simulation but excluding the contribution from the dense
wake behind the shock (see
the lower sequence of points in the bottom frame of Fig. 6). This force
is comparable to the one obtained from the analytic density
stratification given by Eq.~(\ref{rho}), which only refers to the
material in the free-streaming region before entry into the axial wake.
A comparison between the gravitational force with (upper sequence
of points) and without (lower sequence) the material within the
wake indicates that $\sim 90$\%\ of the gravitational force
on the body comes from the material within the wake.

\section{Summary}

We present an analytic model for the flow generated by a point mass 
moving hypersonically within a homogeneous environment. This model
is based on the ballistic orbit theory (see, e.~g., Bisnovatyi-Kogan
et al.~1979), and is developed so as to obtain analytic expressions
for the mass accretion rate and the non-linear gravitational drag. Since we
include the contribution of the non-linear inner wake, 
there is no ambiguity in the definition of the minimum cut-off distance
of the interaction, which turns out to be $\simeq 0.82\xi_{0}$.

We find that the predicted mass accretion rate and gravitational drag
agree satisfactorily with the results from an axismmetric, isothermal
simulation:
\begin{itemize}
\item For the mass accretion rate we essentially find full
agreement between the analytic and numerical results (see Eq. \ref{m} and
Fig. 6). This result in principle differs from the one of Moekel \&
Throop (2009), who obtain a significantly lower value for the
accretion rate from their numerical simulation. The fact that
we obtain a better agreement could be due to the considerably
higher resolution of our simulation, or to the fact that we carry
out a much longer time-integration (extending to $\sim 70\xi_0/v_0$,
compared to $\sim 1.5 \xi_0/v_0$ for the simulation of Moekel
\& Throop 2009).
\item For the net drag force $F_d$ we obtain an agreement within
$\sim 20$~\%\ between the prediction from the analytic model
and the numerical simulations (see Fig. 6).
Though the analytic and numerical drag forces have a reasonable
quantitative agreement, it appears that the analytic model predicts
a shallower $F_d$ vs. $R$ dependence (where $R$ is the radius of the
control volume enclosing the material assumed to produce
the drag) than the one obtained from the numerical simulation
(see Fig. 6).
\end{itemize}
There are several possible sources for this discrepancy between
the analytic and numerical drag forces. It appears that the limited
numerical resolution of the simulation is not responsible for this
effect, because we have repeated the simulation at $1/2$ and $1/4$ of
the resolution (of the simulation presented in section 4) and
obtain basically the same drag force. A possible source of the
differences between the analytic and numerical $F_d$ is the
fact that the simulation has a finite Mach number (${\mathcal{M}}=5$ for the
upstream flow, see section 4), while the analytic model essentially
has an infinite Mach number (i.e., zero gas pressure). Another
difference is that the numerical simulation has a rather broad
wake region, while in the analytic solution it is assumed that
the tail occupies a very narrow region surrounding the symmetry axis.
A third difference is that while in the analytic model the perturbed
environmental gas effectively extends to infinity, the numerical
simulation of course is carried out in a finite domain (see section 4).
Given these clear differences between the numerical and analytic models,
the agreement that we find between the two can be regarded as quite
successful.

In this paper we have therefore derived an analytic recipe for the
drag force $F_d$ (from a ballistic flow model), which is
successfully reproduced by an axisymmetric numerical simulation.
This recipe for $F_d$ will be useful for carrying out simulations
of compact bodies in motions influenced by gravitational drag. Possible
examples are the motions of young stars within molecular clouds
(see, e.~g., Throop \& Bally 2008; Chavarr\'{\i}a et al.~2010), or the orbital decay of black holes
in the centre of merging galaxies (Narayan 2000; Escala et al.~2004, 2005;
Dotti et al.~2006). Kim \& Kim (2009) computed the nonlinear gravitational 
drag on a massive Plummer perturber in adiabatic axisymmetric 
simulations and  found that it is smaller than the linear theory predicts for
supersonic bodies.
This reduction of the drag force is accounted for correctly in our drag formula.

\section*{Acknowledgments}
We acknowledge support from the CONACyT grants 60526,
61547, 101356 and 101975.

\bsp

\label{lastpage}

\end{document}